\documentclass{iaus}
\usepackage{graphicx}

\title[Depth of the Magellanic Clouds] 
{Line of sight depth of the Large and Small Magellanic Clouds}

\author[Annapurni Subramaniam \& Smitha Subramanian]   
{Annapurni Subramaniam
 \and Smitha Subramanian}

\affiliation{Indian Institute of Astrophysics \\ Sarjapur Road,
Koramangala II Block, Bangalore -560034 \\ email: {\tt purni@iiap.res.in, smitha@iiap.res.in}} 

\pubyear{2008}
\volume{256}  
\pagerange{119--126}
\jname{The Magellanic System: Stars, Gas, and Galaxies}
\editors{Jacco Th. van Loon \& Joana M. Oliveira, eds.}
\begin{document}

\maketitle
\begin{abstract}
We used the red clump stars from the Optical Gravitational Lensing Experiment  (OGLE II) survey and the 
the Magellanic Cloud Photometric Survey (MCPS), to estimate the line of sight depth. 
The observed dispersion in the magnitude and colour distribution of red clump stars is used to
estimate the line of sight depth, after correcting for the contribution due to
other effects. This 
dispersion due to depth, has a range from minimum dispersion that can be estimated, to 0.46 mag
(a depth of 500 pc to 10.44 Kpc), in the LMC. In the case of SMC, the dispersion ranges
from minimum dispersion to 0.35 magnitude (a depth of 665 pc to 9.53 Kpc).
The thickness profile of LMC bar indicates that it is flared. The average depth in the bar region
is 4.0$\pm$1.4 kpc.  The halo of the LMC (using RR Lyrea stars) is found to have larger depth
compared to the disk/bar, which supports the presence of inner halo for the LMC.
The large depth estimated for the LMC bar and the disk suggests that the LMC might have had minor
mergers. In the case of SMC, the bar depth (4.90$\pm$1.23 Kpc) and the disk depth (4.23$\pm$1.48 Kpc)
are found to be within the standard deviations. 
We find evidence for increase in depth near the optical center (up to 9 kpc).
On the other hand, the estimated depth for the halo (RR Lyrea stars) and disk (RC stars)
for the bar region of the SMC is found to be similar. Thus, increased depth
and enhanced stellar as well as HI density near the
optical center suggests that the SMC may have a bulge.

\end{abstract}

\firstsection 
\section{Introduction}
The Magellanic Clouds were believed to have interactions with our Galaxy as well as between each other
\cite[(Westerlund 1997)]{west97}. 
The N-body simulations by \cite[Weinberg (2000)]{wein00} predicted that the LMC's evolution is
significantly affected by its interactions with Milky Way and the tidal forces will thicken
and warp the LMC disk. \cite[Alves and Nelson (2000)]{an00}
 studied the carbon star kinematics and found that
the scale height, h, increases from 0.3 to 1.6 kpc over the range of radial distance, R,
from 0.5 to 5.6 kpc and hence concluded that the LMC disk is flared. Using an expanded sample
of carbon stars \cite[van der Marel et al. (2002)]{vdm02}
 also found that the thickness of LMC disk increases
with the radius. There has not been any direct estimate of the thickness or the line of sight
depth of the bar and disk of the L\&SMC so far.

\cite[Mathewson, Ford \& Visvanathan (1986)]{mfv86} found that SMC Cepheids extend from
43 to 75 Kpc with most Cepheids found in the neighbourhood of 59 Kpc.
Later, the line of sight depth of SMC was
estimated \cite[(Welch et al. 1987)]{welch87} by investigating the line of sight distribution and
period - luminosity relation of Cepheids and
found the line of sight depth of SMC to be around 3.3 Kpc. 
\cite[Hawkins et al. (1989)]{haw89},
estimated the line of sight depth in the outer regions of SMC to be around 10-20 kpc.
Measurements of thickness in the central regions of Magellanic Clouds,
especially LMC, is of strong interest to understand the contribution of LMC's
self lensing to the observed microlensing events from this Galaxy.\\

Red Clump (RC) stars are core helium burning stars which are metal rich and slightly
more massive counter parts of
horizontal branch stars.  In the this paper, we use the dispersions in the
colour and magnitude distribution of RC stars for the depth estimation. The dispersion in
colour is due to a combination of observational error, internal reddening (reddening within the disk
of the LMC/SMC) and population effects. The dispersion in magnitude is due to internal
disk extinction, depth of the distribution, population effects and photometric errors
associated with the observations. By deconvolving other effects from the dispersion of
magnitude, we can estimate the depth of the disk. The advantage of choosing RC stars
as proxy is that there are large number of these stars available to determine the
dispersions in their distributions with good statistics.

\section{Data}
Data for the LMC is taken from the OGLE II catalogue \cite[(Udalski \etal\ 2000)]{u00}
The average photometric error
of red clump stars in I and V bands are around 0.05 magnitude. Photometric data with error less
than 0.15 mag are considered for the analysis.  The 26 strips of LMC are divided into 1664 regions.
(V$-$I) vs I CMDs are plotted for each region and red clump stars ($\ge$ 1000 stars) were identified.
For all the regions, red clump stars are well within a box in CMD, with boundaries
0.65 - 1.35 mag in (V$-$I)colour and 17.5 - 19.5 mag in I magnitude. The OGLE data
suffers from incompleteness due to crowding effects and it is corrected using the
data given in \cite[Udalski \etal\ (2000)]{u00}.
 
The data for the SMC is taken from two surveys (OGLE II: \cite[Udalski \etal\ 1998]{u98} \&
MCPS: \cite[Zaritsky \etal\ 2002]{z02}). The OGLE \&\ MCPS photometric
data with error less than 0.15 mag are considered for the analysis. The observed regions
of SMC OGLE and MCPS data are divided into 176 and 876 sub regions respectively.In the MCPS data the
regions away from the bar are less dense compared to the bar region and out of 876 regions only 
755 regions which have reasonable number of stars ($\sim$ 1000 stars) are considered for the analysis. 
For all the regions, red clump stars are identified in the same region in the 
CMD with boundaries 0.65 - 1.35
mag in (V$-$I) colour and 17.5 - 19.5 mag in I magnitude. OGLE data is corrected for incompleteness.

\section{Analysis}
A spread in magnitude and colour of red clump stars is observed in the CMDs of both 
LMC and SMC. The number distribution profile against colour and magnitude roughly 
resembles a Gaussian. The width of the Gaussian in the distribution of colour and magnitude is obtained
using a non linear least square method to fit the profile.

The RC stars in the disk of LMC/SMC is a heterogeneous population and the stars have a range
in mass, age and metallicity. The density of stars in various location will also vary
with star formation rate as a function of time. \cite[Girardi \& Salaris (2001)]{g01} simulated the RC 
stars in the LMC using the star formation rate results 
from \cite[Holtzman \etal\ (1999)]{h99}
 and the age metallicity relation from 
\cite[Pagel and Tautvaisiene (1998)]{pt98}. They also simulated the RC 
stars in SMC using star formation results and age metallicity relation from 
\cite[Pagel and Tautvaisiene (1998)]{pt98}.
The intrinsic dispersions obtained from the above model is used in our estimation to account
for the population effects. The estimated intrinsic dispersions in magnitude and colour distributions
according to the theoretical model for LMC are 0.1 and 0.025 mag respectively. In the case of
SMC, the values are 0.076 and 0.03 mag respectively.

The average photometric errors of I and V band magnitudes are calculated for each region
and the error in I and (V$-$I) colour are estimated. These are subtracted from the observed
width of magnitude and colour distribution respectively.
After correcting for population effects and the observational error in colour, the remaining
spread in colour distribution is taken as due to the internal reddening, E(V$-$I) . This is
converted into extinction in I band using the relation A(I) = 0.934 E(V$-$I), where E(V$-$I) is the
internal reddening value estimated for each location. The above relation is derived from
the relations E(V$-$I) = 1.6 E(B$-$V) and A(I) = 0.482 A(V) \cite[(Rieke and Lebofsky 1985)]{rl85}.
 
The following relations are used for estimating the resultant dispersion due to depth.

$\sigma$$^2$$_m$$_a$$_g$ = $\sigma$$^2$$_d$$_e$$_p$$_t$$_h$ +
$\sigma$$^2$$_i$$_n$$_t$ $_e$$_x$$_t$$_i$$_n$$_c$$_t$$_i$$_o$$_n$ +
$\sigma$$^2$$_i$$_n$$_t$$_r$$_i$$_n$$_s$$_i$$_c$ + $\sigma$$^2$$_e$$_r$$_r$$_o$$_r$\\
$\sigma$$^2$$_c$$_o$$_l$ =
$\sigma$$^2$$_i$$_n$$_t$$_e$$_r$$_n$$_a$$_l$ $_r$$_e$$_d$$_d$$_e$$_n$$_i$$_n$$_g$ +
$\sigma$$^2$$_i$$_n$$_t$$_r$$_i$$_n$$_s$$_i$$_c$ + $\sigma$$^2$$_e$$_r$$_r$$_o$$_r$

\begin{figure}[b]
\begin{center}
\includegraphics[width=5.0in]{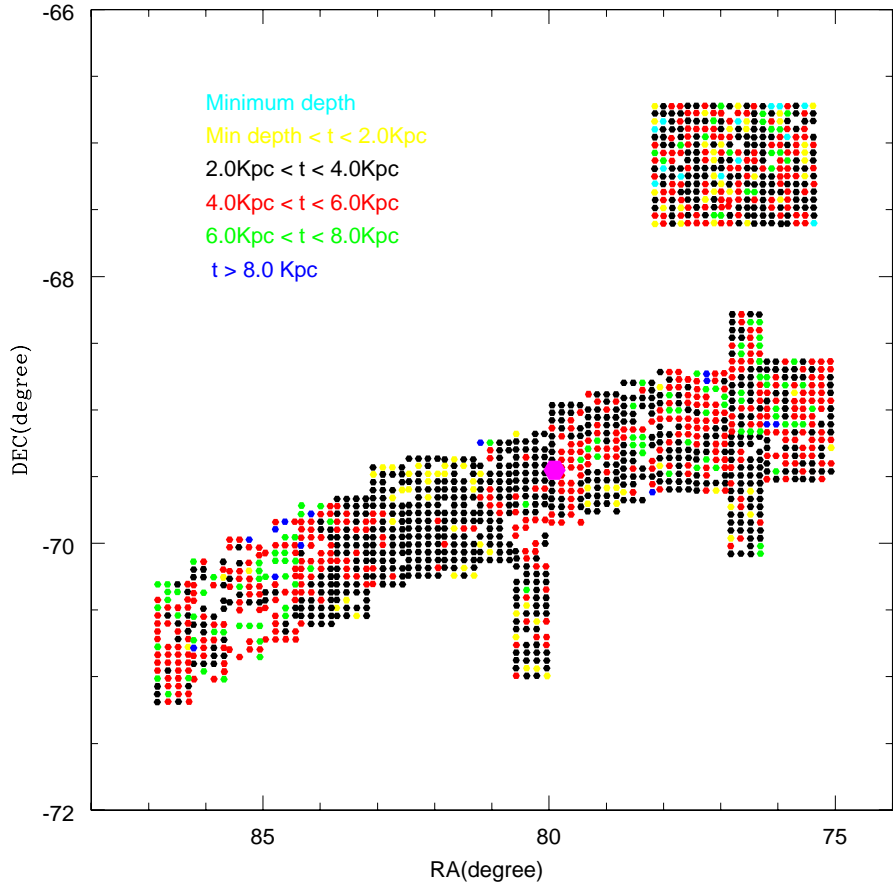} 
\caption{The line of sight depth in the bar region of the LMC. The colour codes
used to denote the depth is explained. The optical center is shown in purple dot.}
\label{fig1}
\end{center}
\end{figure}

\begin{figure}[b]
\begin{center}
\includegraphics[width=5.0in]{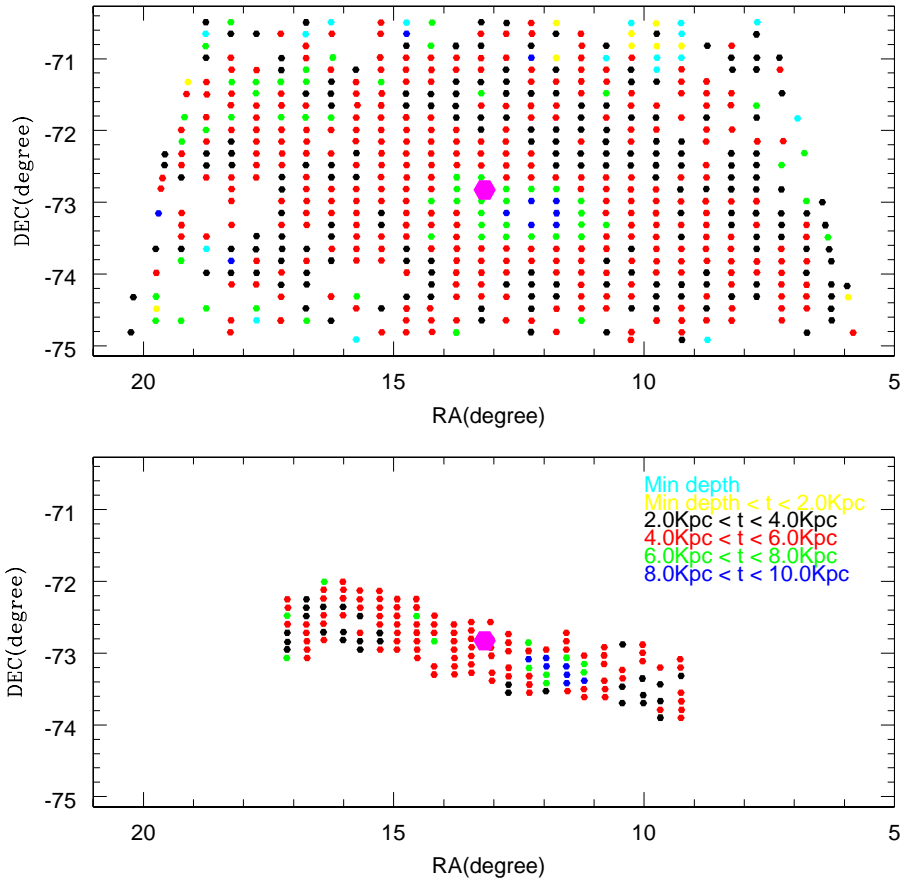} 
\caption{The line of sight depth in the bar and disk region of the SMC. The upper
panel shows the depth derived from the MCPS data and the lower panel shows that derived from
 the OGLE data. The optical center is shown in purple dot.}
\label{fig2}
\end{center}
\end{figure}

\section{Results: LMC}
A two dimensional plot of the depth for the 1528 regions in the LMC is shown in figure 1.
The optical center of LMC is taken to be RA = 05$^h$ 19$^m$ 38$^s$,
Dec = $-$69$^0$ 27' 5".2 (2000, \cite[de Vaucoulers and Freeman,1973]{dvf73}). The plot
shows a range of dispersion values from 0.033 to 0.46 mag (a depth of
of 700 pc to 10.5 kpc;  avg: 3.95 $\pm$ 1.42 kpc) for LMC central bar region. 
For the N-W disk region,
the dispersion estimated ranges from minimal dispersion that
can be estimated (limited by errors), to 0.33 mag (a depth of 500 pc to 7.7 kpc;
avg: 3.56 $\pm$ 1.04 kpc).  Regions in the bar between
RA 80 - 84 degrees show a reduced depth (0.5 - 4 kpc, as indicated by yellow and black points).
The regions to the east and west of the above region are found to have larger depth (2.0 - 8.0 kpc,
black, red and green points). Thus, the depth of the bar at its ends are larger than that near
its center. The  N-W region has depth similar to the central region of the bar.
In general, thicker and heated up bar could be considered as signatures of minor mergers. Thus, the
LMC is likely to have experienced minor mergers in its history.

\cite[Subramaniam (2006)]{s06} studied the distribution of RR Lyrae in the bar 
region of the LMC. She found that
the RR Lyrae stars in the bar region have a disk like distribution, but halo like location.
The RR Lyrea stars are in the same evolutionary state as the RC stars, except that the RR Lyrea
stars belong to an older and metal poor population and hence a proxy for the halo as it belongs
to the Population II stars. We used the total depth estimated for RR Lyrae stars
 and compared it with the
RC depth which can be considered as proxies for halo and disk respectively. 
It was seen that the depth estimated from RR Lyrae stars 
ranged between 4.0 -- 8.0 kpc, suggesting that the RR Lyrae stars span a larger depth
than the RC stars. Thus, at least in the central region of the LMC, the halo, as delineated by
the RR Lyrae stars has much larger
depth than the disk, as delineated by the RC stars. This supports the idea that there is an
inner halo for the LMC. 

\begin{figure}[b]
\begin{center}
\includegraphics[width=4.5in]{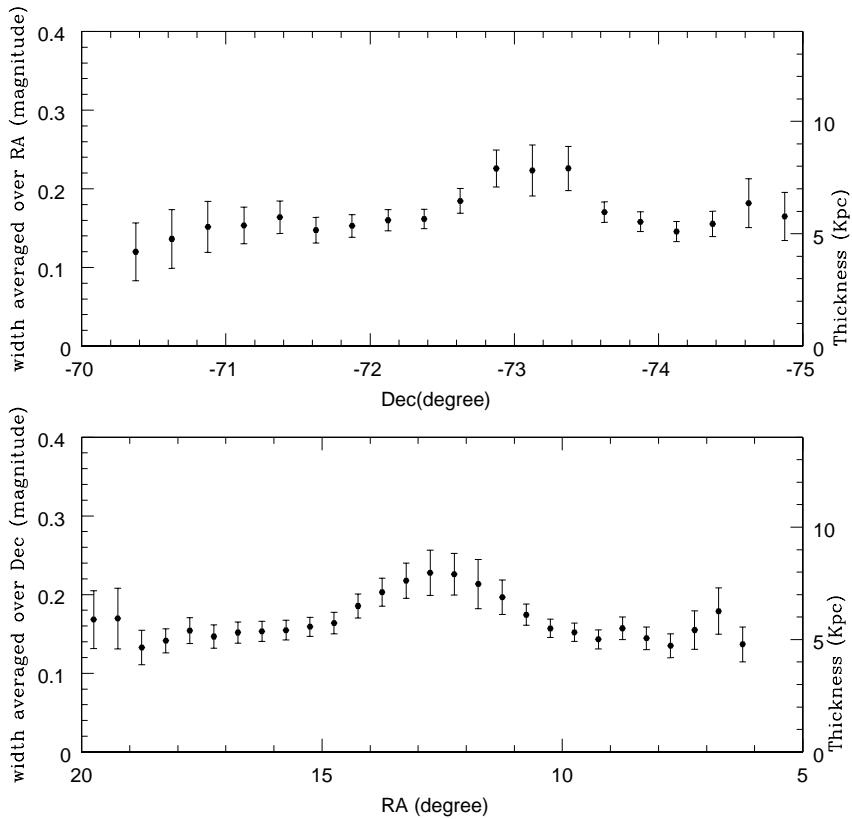} 
\caption{The line of sight depth averaged over a narrow range of RA and Dec near the center using
the MCPS data.
The errors correspond to the the standard deviation over the area averaged.}
\label{fig3}
\end{center}
\end{figure}

\section{Results: SMC}
A colour coded two dimensional
plot of thickness for the two data sets are shown in figure 2 (OGLE data in the lower panel
and MCPS data in the upper panel). The optical center of SMC is
taken to be RA = 00$^h$ 52$^m$ 12.5$^s$ , Dec = $-$72$^0$ 49' 43" 
(2000, \cite[de Vaucoulers and Freeman, 1973]{dvf73}).  The prominent feature in both the plots is the
presence of blue and green points indicating increased depth, for regions located
near the SMC optical center ($\sim$ 9kpc). 
The net dispersions range from 0.10 to 0.35 mag (corresponding to a depth of 2.8 kpc to 9.6 kpc) in
OGLE data set and from minimum dispersion to 0.34 (corresponding to a depth of 665 pc to 9.47 Kpc)
in the MCPS data set. The average value of SMC thickness estimated using OGLE data
set in the central bar region is 4.9$\pm$ 1.2 kpc and
the average thickness estimated using MCPS data set which covers a larger area than
OGLE data is 4.42 $\pm$ 1.46 kpc. The average depth obtained for the bar region alone
is 4.97 $\pm$1.28 kpc, which is very similar to the value obtained from OGLE data. The depth
estimated for the disk alone is 4.23$\pm$1.47 kpc. Thus the disk and the bar of the SMC do
not show any significant difference in the depth. Evidence of increased depth in the outer regions
of the disk is also indicated, this might suggest large depth for the outer SMC.
The enhanced central depth is shown in the averaged width over a limited RA and Dec in figure 3,
using the MCPS data. The increased
depth near the center is clearly indicated. Thus, the depth near the center is about 9.6 kpc, which
is twice the average depth of the bar region (4.9 kpc). 
The depth profile, especially the one along RA, is very similar to the 
luminosity profile of bulges and hence suggests the presence of a bulge in the SMC.

A comparison can be made between the halo and disk/bar of the SMC. The RR Lyrae
stars from the OGLE II data were analysed similar to the procedure adopted by
\cite[Subramaniam (2006)]{s06} and the dispersion due to depth alone was estimated. 
In contrary to what is
seen in the case of the LMC, both the population show very similar dispersion in the SMC.
The comparison not only suggests that the RR Lyrea stars and the RC stars occupy similar depth, but
also indicates that they show similar depth profile across the bar. The increased depth near the
optical center is also closely matched. 
This suggests that the RR Lyrea stars and RC stars are born in the same location
and occupy similar volume in the galaxy.  This co-existence
and the similar depth of RR Lyrae stars and the RC stars in the central region of the SMC
can be easily explained, if it is the bulge. 
We also find that the old and young stellar density as well as the HI density show
peaks near the region with bulge like depth. This supports the idea that the central region of
the SMC could be its bulge. The elongation and the rather non-spherical appearance of the
bulge could be due to tidal effects or minor mergers \cite[(Bekki \& Chiba 2008)]{bc08}.

\end{document}